\documentclass[twocolumn,superscriptaddress,showpacs,prb,altaffilletter,prl]{revtex4}

\usepackage{graphicx}    
\usepackage{dcolumn}     
\usepackage{bm}          
\usepackage{amsmath}
\usepackage{amssymb}
\usepackage{times}

\begin{document}

\title{Model for vortex-core tunneling spectroscopy
of chiral $p$-wave superconductors
via odd-frequency pairing states }

\author{Yasunari Tanuma}
\affiliation{Faculty of Engineering and Resource Science,
Akita University, Akita 010-8502, Japan}
\author{Nobuhiko Hayashi}
\affiliation{CCSE, Japan Atomic Energy Agency, and CREST-JST, 
6-9-3 Higashi-Ueno, Tokyo 110-0015, Japan}
\author{Yukio Tanaka}
\affiliation{Department of Applied Physics
and CREST-JST, Nagoya University, Nagoya 464-8603, Japan}
\author{Alexander A. Golubov}
\affiliation{Faculty of Science and Technology, University of Twente, 7500 AE, Enshede,
The Netherlands}
\date{\today}

\begin{abstract}
The local density of states (LDOS) is studied theoretically
in terms of the odd-frequency (odd-$\omega$) Cooper pairing induced
around a vortex core.
%
%
We find that a zero energy peak
in the LDOS at the vortex center is robust against nonmagnetic impurities
in a chiral $p$-wave superconductor
owing to an odd-$\omega$ $s$-wave pair amplitude.
We suggest how to discriminate a spin-triplet pairing symmetry
and spatial chiral-domain structure
by scanning tunneling spectroscopy via odd-$\omega$ pair amplitudes inside vortex cores.
\end{abstract}

\pacs{
74.20.Rp,  
74.70.Pq,  
74.50.+r   
}
\maketitle

The unambiguous determination of Cooper pairing symmetry
is of prime importance for understanding the pairing mechanism
of unconventional superconductivity.
In the last decade, the discovery of the ruthenate superconductor (SC)
$\mathrm{Sr}_2\mathrm{RuO}_4$
has stimulated an enormous amount of studies,
where
a chiral superconductivity with time-reversal-symmetry breaking
was indicated by muon-spin-rotation and polar Kerr effect experiments,
and a spin-triplet state such as a $p$-wave one was suggested
by Knight-shift measurements \cite{MaenoRMP,Xia}.
Therefore, a spin-triplet chiral (e.g., $p_x \pm \mathrm{i}p_y$)
state has been considered as a most promising pairing state
\cite{Sigrist}.
However, in order to confirm the chiral $p$-wave
state definitely,
much more clear experiments based on new ideas are needed \cite{Asano07}.
%
%
%
\par
One way of identifying the pairing symmetry among possible candidates
is to resolve the quasiparticle states, by tunneling spectroscopy,
via the surface Andreev bound states (SABSs) \cite{TKTS01}.
The SABS \cite{Hu} originates from a sign change of anisotropic pair potentials
at an interface,
and it is observed as a zero-bias conductance peak
in various materials \cite{TKTS01}.
It is also well known that the so-called vortex Andreev bound states (VABS) 
are formed around a vortex core \cite{Hess}. These bound states can be interpreted 
as the odd-frequency (odd-$\omega$) pairing states \cite{YTG}.
The odd-$\omega$ pairing state is characterized by a pair amplitude
that is an odd function of the Matsubara frequency \cite{Efetov}.
The origin of the generation of
odd-$\omega$ pair amplitude is as follows.
In inhomogeneous system, due to the breakdown
of translational invariance, the pair potential acquires
a spatial dependence which leads to coupling between the
even- and odd-parity pairing states.
The Fermi-Dirac statistics
then dictates that the pair amplitude of opposite parity should be opposite in frequency \cite{TG07}.
%
%
Owing to an axial symmetry of a vortex,
there exists a fundamental rule \cite{Salomaa,Kato01} that relates
the angular momentum of the odd-$\omega$ Cooper pair
and the topology (winding number) of vortex.
It is interesting and important to confirm experimentally the existence of
odd-$\omega$ pair amplitudes subjected to the topological symmetry rule
due to the vortex.
Note that the SABSs are also interpreted as a generation of
the odd-$\omega$ pairing states \cite{TG07,TTG07}.
%
\par
%
%
In this Letter,
we will show that nonmagnetic impurity scattering effect on the VABS
can be used to detect the symmetry rule for
the induced odd-$\omega$ pair amplitudes.
We will study the odd-$\omega$ pairing amplitude and
the local density of states (LDOS)
around a vortex for spin-singlet $s$-wave and spin-triplet chiral $p$-wave SCs.
The effect of nonmagnetic impurities serves as
a probe of the symmetry rule and the pairing symmetry
in candidates for a spin-triplet chiral SC such as Sr$_2$RuO$_4$.
We will also point out how to observe spatial chiral-domain structure
by scanning tunneling spectroscopy/microscopy (STS/STM).
%
The issue of chiral domains has been studied experimentally
in Sr$_{2}$RuO$_{4}$ \cite{Mota,Kirtley}.
However, a more direct observation by 
spatially resolved probe has not been performed. 
Therefore, it is interesting to propose an idea to
observe directly the spatial structure of
chiral domains.
%
In the present theory, we point out that
the proposed method for identifying a spin-triplet chiral $p$-wave state
can be simultaneously used for detecting chiral domains.
\par
We consider a spin-singlet $s$-wave and a spin-triplet chiral $p$-wave pairing state
with a $\bm{d}$-vector parallel to the $z$ axis \cite{MaenoRMP,Sigrist}.
In both cases, the quasiclassical
Green's function $\hat{g}(\mathrm{i}\omega_n, \bm{r}, \bar{\bm{k}})$
is represented in a $2 \times 2$ matrix form as \cite{HayaLT}
\begin{align}
\hat{g}
= -\mathrm{i}\pi \left (
\begin{array}{cc}
g &
\mathrm{i} f \\
-\mathrm{i} {\bar f} &
- g
\end{array}
\right ).
\end{align}
It follows the Eilenberger equation
\begin{align}
\label{eq:Eilen}
-\mathrm{i} {\bm v}_{\rm F} \cdot \nabla \hat{g}
= \left [ \mathrm{i}\omega_n \hat{\tau}_z
-\hat{\Delta} - \hat{\Sigma} , \hat{g}
\right ],
\end{align}
%
which is supplemented by the normalization condition $g^{2} + f \bar{f} =1$.
Here, $\omega_n = (2n+1)\pi T$ is the Matsubara frequency,
${\bm v}_{\rm F}$ is the Fermi velocity,
%
and
$\hat{\bm{\tau}}=(\hat{\tau}_x,\hat{\tau}_y,\hat{\tau}_z)$
are Pauli matrices in the particle-hole space.
%
We use units in which $\hbar=k_{\rm B}=1$.
$\bm{r}$
denotes the center of mass of the Cooper pair,
and $\bar{\bm{k}}$ is the unit vector of the Fermi wave number
($\bar{\bm{k}}={\bm k}_{\rm F}/|{\bm k}_{\rm F}|$).
%
%
%
%
The pair potential is expressed as
$
\hat{\Delta}(\bm{r}, \bar{\bm{k}})
=
({\hat \tau}_x+\mathrm{i}{\hat \tau}_y)\Delta (\bm{r},\bar{\bm{k}})/2
-({\hat \tau}_x-\mathrm{i}{\hat \tau}_y)\Delta^{*}(\bm{r},\bar{\bm{k}})/2
$.
%
%
We incorporate the impurity scattering effect
within the Born approximation.
%
The impurity self energy
is given as \cite{HayaLT}
$
\hat{\Sigma}(\mathrm{i}\omega_n,\bm{r})
=\Gamma
\bigl[
({\hat \tau}_x+\mathrm{i}{\hat \tau}_y)
\bigl\langle f(\mathrm{i}\omega_n, \bm{r}, \bar{\bm{k}}) \bigr\rangle
-
({\hat \tau}_x-\mathrm{i}{\hat \tau}_y)
\bigl\langle {\bar f}(\mathrm{i}\omega_n, \bm{r}, \bar{\bm{k}}) \bigr\rangle
-2\mathrm{i}{\hat \tau}_z
\bigl\langle g(\mathrm{i}\omega_n, \bm{r}, \bar{\bm{k}}) \bigr\rangle
\bigr]/2
$,
%
%
where the brackets $\langle \cdots \rangle$ denote the average over
the Fermi surface.
We define the impurity scattering rate in the normal state
as $\Gamma =1/2\tau$ with mean free path
$v_{\rm F}\tau $.
\par
We assume an isotropic two-dimensional system
and
introduce the angle $\theta$ as
$\bar{\bm{k}} = (\cos\theta, \sin\theta)$.
The chiral $p_x \pm \mathrm{i}p_y$ ($=p_\pm$) wave is represented
by $\exp(\pm \mathrm{i}\theta)$.
The self-consistent equation for the pair potential
is given by
\begin{align}
\label{eq:PP}
\Delta (\bm{r},\theta)
&=\Delta_{+}(\bm{r})e^{+\mathrm{i}l\theta}
+\Delta_{-}(\bm{r})e^{-\mathrm{i}l\theta},
\\
\Delta_{\pm} (\bm{r})
&=\pi T V \sum_{|\omega_n| < \omega_c}
\Bigl\langle
e^{\mp\mathrm{i}l\theta^{\prime}}
f(\mathrm{i}\omega_n,\bm{r},\theta^{\prime})
\Bigr\rangle,
\end{align}
where $V$ is the coupling constant (see Ref.\ \cite{HayaLT} for details).
%
%
%
We consider
a spin-singlet $s$-wave state ($l=0$)
and
a spin-triplet chiral $p$-wave one ($l=1$)
for the pair potential.
The Fermi-surface average is
$\langle \cdots \rangle=\int^{2\pi}_{0} d\theta^{\prime} \cdots /(2\pi)$.
The pair amplitudes represented in the Matsubara frequency are
\begin{align}
\label{eq:PA}
F^{(l)}(\mathrm{i}\omega_n,\bm{r},\theta)&=
F^{(l)}_{+}(\mathrm{i}\omega_n,\bm{r})e^{+\mathrm{i}l\theta}
+F^{(l)}_{-}(\mathrm{i}\omega_n,\bm{r})e^{-\mathrm{i}l\theta},
\\
F^{(l)}_{\pm}(\mathrm{i}\omega_n,\bm{r})
&=
\Bigl\langle
e^{\mp\mathrm{i}l\theta^{\prime}}
f(\mathrm{i}\omega_n,\bm{r},\theta^{\prime})
\Bigr\rangle,
\end{align}
with the quantum number of the
angular momentum $l=0$, 1, 2, $\cdots$,
where the even-$\omega$ and odd-$\omega$
pair amplitudes satisfy
$f(\mathrm{i}\omega_n,\bm{r},\theta^{\prime})
=f(-\mathrm{i}\omega_n,\bm{r},\theta^{\prime})$,
and
$f(\mathrm{i}\omega_n,\bm{r},\theta^{\prime})
=-f(-\mathrm{i}\omega_n,\bm{r},\theta^{\prime})$,
respectively.
\par
The Eilenberger equation
is simplified
by introducing the Riccati parameterization \cite{Schopohl}.
%
%
%
%
%
%
%
We numerically solve the resulting Riccati equations
and the self-consistent equations for the impurity self energy
and pair potential iteratively
\cite{HayaLT}.
Throughout the paper
we set the temperature $T=0.1T_{\rm c}$,
where $T_{\rm c}$ is the critical temperature in the absence of impurities.
Using the self-consistently obtained pair potential,
we determine the self energy by the analytical continuation
with  $\mathrm{i}\omega_n \rightarrow E+\mathrm{i}\delta$ \cite{Kato02}.
%
The LDOS is then calculated as
$
N({\bm r},E) = N_{\rm F}
\bigl\langle
\mathrm{Re} \ g^{\rm R}
\bigr\rangle
$.
Here,
$N_{\rm F}$ is the normal-state density of states
at the Fermi level,
$g^{\rm R}=g(\mathrm{i}\omega_n \rightarrow E+\mathrm{i}\delta)$,
$E$ means the quasiparticle energy, and $\delta$
is an infinitesimal quantity.
We select $\delta = 0.06 \Delta_0$ as a typical value,
where $\Delta_0$ is the bulk amplitude of the pair potential at $T=0$ and $\Gamma=0$.
\par
An enhancement of the LDOS in the presence of
odd-$\omega$ pairing can be understood
by means of the normalization condition.
Since $\bar{f}$ with
$E=0$ is given by $\bar{f}=-f^{*}$ for odd-$\omega$ pairing state \cite{YTG},
one can show that generally 
$N(E=0)/N_{\rm F}>1$ owing to $g^{2}=1 + |f|^{2} > 1$. 
This means that the emergence of odd-$\omega$ pairing
is a physical reason for a zero energy peak (ZEP) in the LDOS. 
\par
%
\begin{figure}[htb]
\begin{center}
\scalebox{0.40}{
\includegraphics{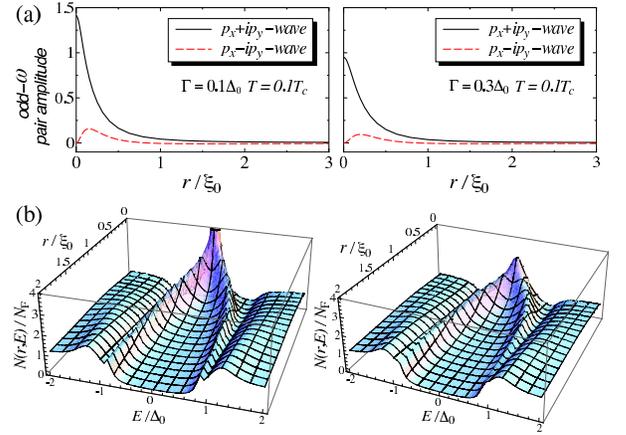}
}
\caption{(Color online)
(a) Spatial dependence of the odd-$\omega$ pair amplitudes ($\omega_{n = 0}$),
and (b) the corresponding LDOS
for the $s$-wave vortex
with $\Gamma =0.1\Delta_0$ (left panel) and $0.3\Delta_0$ (right panel).
%
%
%
\label{fig:01}}
\end{center}
\end{figure}
%
In circular coordinates ${\bm r}=(r,\phi)$,
we consider an axially-symmetric vortex situated at ${\bm r}=0$,
the vorticity of which
is perpendicular to the two-dimensional system.
The axial symmetry requires
$\Delta \to \Delta \exp(iN \alpha)$
with an integer $N$
for a rotational transformation $\phi \to \phi + \alpha$
and $\theta \to \theta + \alpha$ \cite{Salomaa,Kato01,HayaLT}.
The pair potential that satisfies this constraint is
\begin{align}
\Delta (\bm{r},\theta)
&={\bar \Delta}(r)e^{\mathrm{i}m\phi},
\end{align}
for a spin-singlet $s$-wave vortex, and
\begin{align}
\Delta (\bm{r},\theta)
&={\bar \Delta}_{+}(r)e^{\mathrm{i}[\theta+(m-2)\phi]}
+{\bar \Delta_{-}}(r)e^{\mathrm{i}[-\theta+m\phi]},
\end{align}
for a spin-triplet chiral $p$-wave vortex.
Here, $m$ is the winding number of a vortex.
We have assumed the $p_-$-wave state
in bulk for the chiral $p$-wave vortex,
namely
${\bar \Delta}_{+}(r \to \infty)=0$
and
${\bar \Delta}_{-}(r \to \infty) \neq 0$.
Due to the same constraint from axial symmetry,
the pair amplitudes satisfy
\begin{align}
\label{eq:amp-s}
F^{(l)}_{\pm}(\mathrm{i}\omega_n,\bm{r})
&=
{\bar F}^{(l)}_{\pm}(\mathrm{i}\omega_n,r)
e^{\mathrm{i} (m \mp l) \phi},
\end{align}
for the spin-singlet $s$-wave vortex,
and
\begin{align}
\label{eq:amp-p}
F^{(l)}_{\pm}(\mathrm{i}\omega_n,\bm{r})
&=
{\bar F}^{(l)}_{\pm}(\mathrm{i}\omega_n,r)
e^{\mathrm{i} (m-1 \mp l) \phi},
\end{align}
for the spin-triplet chiral $p$-wave ($p_-$-wave) vortex.
Note that
if a pair amplitude has a finite phase proportional to $\phi$,
the amplitude inevitably becomes zero at $r=0$ where $\phi$ is undefined.
\par
First let us discuss the spin-singlet $s$-wave vortex ($l=0$)
in the presence of impurity scattering.
%
%
The spatial dependencies of
odd-$\omega$ pair amplitudes ($\omega_{n = 0}$) are plotted
in Fig.~\ref{fig:01}(a).
There, the distance $r$ from the vortex center is normalized by
the coherence length $\xi_0=v_{\rm F}/\Delta_0$.
%
In the case of winding number $m=1$,
only the $p_{+}$-wave pair amplitude
with odd-$\omega$ is induced at the vortex center.
It is because only the phase
of the $p_{+}$-wave pair amplitude $F^{(l=1)}_+$ 
is zero in Eq.\ (\ref{eq:amp-s}) for $m=1$.
As a result, the LDOS at the vortex center
has the ZEP as shown in Fig.~\ref{fig:01}(b),
reflecting the VABS
that originates from the induced odd-$\omega$
pair amplitude \cite{TG07,YTG}.
%
%
The magnitude of the odd-$\omega$
pair amplitude
is suppressed with the increase of $\Gamma$.
Accordingly, the height of the  ZEP  decreases
with increasing $\Gamma$
[Fig.~\ref{fig:01}(b)].
Actually, a collapse of the ZEP upon
doping impurities in
an $s$-wave SC was observed experimentally \cite{Renner}.
%
%
%
%
Here, the induced odd-$\omega$ pair amplitude is the $p_{+}$-wave
(i.e., non-$s$-wave), and therefore
Anderson's theorem for non-magnetic impurities \cite{Anderson}
is not applicable.
As a result, the odd-$\omega$ pair amplitude, namely the ZEP,
is sensitive to impurity scattering.
%
%
\par
%
\begin{figure}[htb]
\begin{center}
\scalebox{0.29}{
\includegraphics{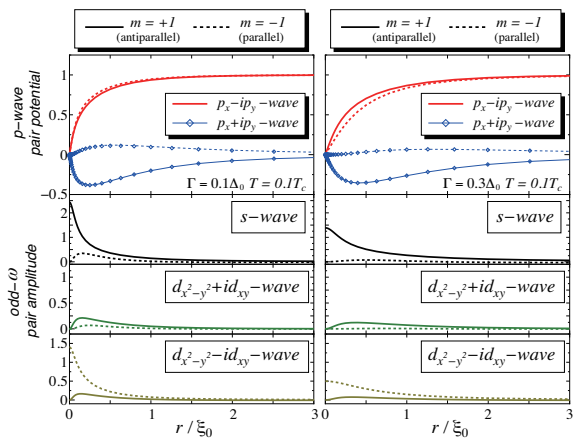}
}
\caption{(Color online)
Spatial dependence of the pair potentials
and the odd-$\omega$ pair amplitudes ($\omega_{n = 0}$)
for the chiral $p$-wave vortices
with $\Gamma=0.1\Delta_0$ (left panel) and $0.3\Delta_0$
(right panel).
The $p$-wave pair potentials are normalized by bulk value of the $p_{-}$-wave one.
%
\label{fig:02}}
\end{center}
\end{figure}

\begin{figure}[htb]
\begin{center}
\scalebox{0.4}{
\includegraphics{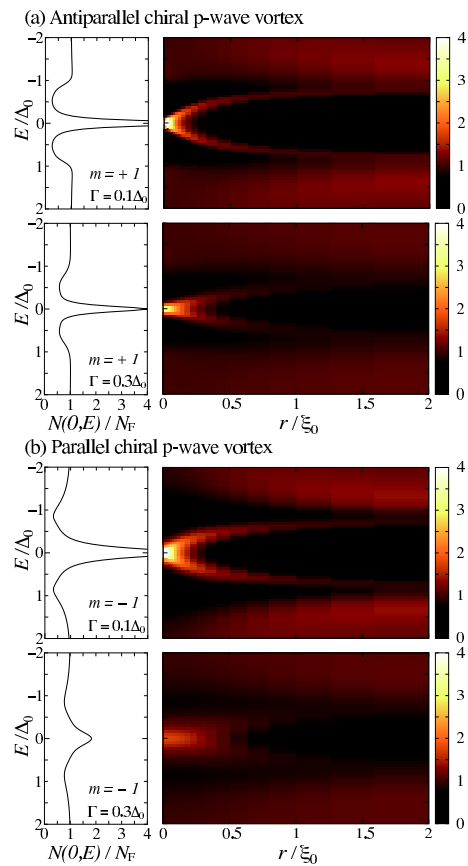}
}
\caption{(Color online)
The LDOS at the vortex center (left)
and the density plot of LDOS
as a function of the distance $r$ (right)
for (a) the antiparallel
and (b) the parallel
vortex.
The upper and lower panels are plots for
$\Gamma = 0.1\Delta_0$ and $0.3\Delta_0$,
respectively.
%
%
\label{fig:03}}
\end{center}
\end{figure}
%
Next, let us move to impurity effects on
the spin-triplet chiral $p$-wave vortices,
the main subject of this paper.
There are two types of vortex,
the vorticity of which is antiparallel ($m=1$) or parallel ($m=-1$)
to the predominant $p_{-}$-wave pair potential \cite{Kato01}.
As seen in Fig.~\ref{fig:02},
the odd-$\omega$ $s$- and
$d_{-}$-wave
pair amplitudes are induced in the vortex center
for the antiparallel and parallel vortices, respectively.
The magnitudes of these odd-$\omega$ pair amplitudes
are reduced with the increase of $\Gamma$.
Note that the decrease of the pair amplitudes with $\Gamma$
is weaker for the antiparallel vortex
than that for the parallel one.
%
As seen from  Figs.~\ref{fig:03}(a) and \ref{fig:03}(b),
while the height of ZEP strongly decreases with $\Gamma$
for the parallel vortex,
it is robust against $\Gamma$
for the antiparallel vortex \cite{Kato02,chiral-p}.
%
%
\par
\begin{figure}[htb]
\begin{center}
\scalebox{0.37}{
\includegraphics{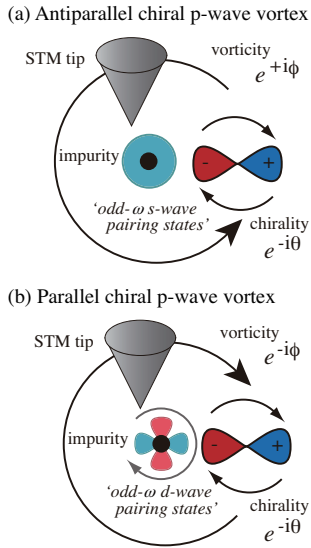}
}
\caption{(Color online)
Schematic illustration of
STS/STM measurements
to detect the odd-$\omega$ pair amplitude
for (a) the antiparallel and (b) the parallel vortex
in chiral $p$-wave SCs.
The arrows represent the phase rotation.
\label{fig:04}}
\end{center}
\end{figure}
\begin{figure}[htb]
\begin{center}
\scalebox{0.31}{
\includegraphics{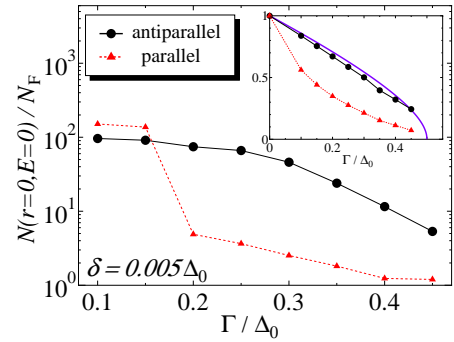}
}
\caption{(Color online)
The ZEP in the LDOS at the vortex center
as a function of $\Gamma$.
%
%
The lines are guides for the eye.
Inset: the odd-$\omega$ pair amplitudes ($\omega_{n=0}=\pi T$)
at the vortex center as functions of $\Gamma$.
The $s$-wave (circle) and $d_{-}$-wave (triangle) pair amplitudes are plotted for
the antiparallel and parallel vortices, respectively (see also Fig.~\ref{fig:02}).
Each pair amplitude is normalized by its value at $\Gamma=0$.
The solid curve represents the Abrikosov-Gorkov law \cite{Mackenzie}.
\label{fig:05}}
\end{center}
\end{figure}

%
%
The clear difference between two states can be
understood in terms of the
symmetry of the odd-$\omega$ pair amplitude.
Due to the phase factor in Eq.~(\ref{eq:amp-p}),
the {\it odd-$\omega$ pair amplitude} induced at the vortex center
is inevitably $s$-wave ($l=0$) for the antiparallel vortex ($m=1$)
and $d_-$-wave ($l=2$) for the parallel one ($m=-1$) as mentioned above.
According to Anderson's theorem \cite{Anderson},
an $s$-wave pair amplitude is robust against non-magnetic impurities,
while a $d_-$-wave pair amplitude is sensitive to such impurities
(see the inset in Fig.~\ref{fig:05}).
The VABS (corresponding to the SABS in Ref.~\cite{TTG07})
originates from an odd-$\omega$ pair amplitude \cite{TTG07},
and it is reflected in the ZEP \cite{YTG}.
Hence, the ZEP is robust against the increase of $\Gamma$
only for the antiparallel vortex where the VABS originates from the $s$-wave pair amplitude.
On the other hand, for the parallel vortex the ZEP is substantially suppressed by $\Gamma$ because of
the $d_-$-wave (i.e., non-$s$-wave) pair amplitude.
Accordingly, measurements of the ZEP under the influence of impurities
correspond to observations of the symmetry of those odd-$\omega$ pair amplitudes
as illustrated in Fig.~\ref{fig:04}.
%
%
In order to visualize the striking difference between the
antiparallel and parallel vortex states, we show in Fig.~\ref{fig:05}
the zero energy LDOS at the vortex center as a function of $\Gamma$.
In the inset of Fig.~\ref{fig:05}, it is found that the odd-$\omega$ $s$-wave pair amplitude
(circle) is likely to follow the Abrikosov-Gorkov (AG) law \cite{Mackenzie},
while the odd-$\omega$ $d_{-}$-wave one (triangle) decays
more rapidly with increasing $\Gamma$.
The decrease of the odd-$\omega$ $s$-wave component is involved with the fact that
the odd-$\omega$ component is generated through the 
coupling to the bulk 
even-$\omega$ $p$-wave pair potential 
which is sensitive to $\Gamma$.
\par
%

%
%
The above results have a significant implication to
identification of pairing symmetry in chiral SCs such as Sr$_2$RuO$_4$.
In chiral superconductivity, the pair potential is composed of two degenerate components.
As a result, degenerate chiral states, such as
$p_x+\mathrm{i}p_y$ and $p_x-\mathrm{i}p_y$, form a domain structure
in a chiral SC
under field cooling condition
with high-speed cooling rate \cite{Mota}.
Under magnetic fields the antiparallel vortex is known to be energetically favorable \cite{Heeb},
and therefore the antiparallel vortex state, namely one of two chiral states,
dominates in a sample under field cooling condition with slow cooling rate.
In the case of {\it high-speed cooling},
a domain structure remains as a mixture of antiparallel- and parallel-vortex domains.
If a SC is a {\it spin-triplet} chiral $p$-wave one,
such a difference in the chiral state between slow and high-speed coolings
is observable via the ZEP at vortex cores shown in Fig.~\ref{fig:04}.
In addition, through a distribution of those two different vortex states,
the existence of a chiral domain structure can be observed
by STM at zero-bias
under high-speed cooling
with a spatial resolution of the order of inter-vortex distance.
On the other hand,
for any spin-singlet pairing state such as a chiral $d$-wave one [e.g., $p_z(p_x\pm \mathrm{i}p_y)$],
the induced odd-$\omega$ pair amplitudes are inevitably spin-singlet {\it odd-parity} ones \cite{YTG}
and the odd-$\omega$ $s$-wave ($=$ {\it even-parity}) pair amplitude is never induced.
%
%
In Sr$_2$RuO$_4$, the ZEP at a vortex core has been observed \cite{Lupien},
and samples with different impurity scattering rate could be prepared \cite{Mackenzie}.
Therefore, an experimental setup proposed here may provide strong
evidence for spin-triplet chiral $p$-wave superconductivity.
%
%
\par
In conclusion,
we have studied the odd-$\omega$ pair amplitudes
and the LDOS around a single vortex
in $s$-wave and chiral $p$-wave SCs.
For the antiparallel and parallel chiral $p$-wave vortices,
we have found that
the odd-$\omega$ $s$-wave and
$d_{-}$-wave pair amplitudes
are inevitably induced at the vortex center, respectively.
The robustness of the ZEP at the vortex core against non-magnetic impurities
originates from this odd-$\omega$ $s$-wave pair amplitude.
Those odd-$\omega$ pair amplitudes
can be observed by STS/STM,
serving as a probe of a spin-triplet pairing
and of a spatial distribution of the chiral domains.
%
%
\par
%
We thank Y. Kato, M. Machida, and S. Kawabata
for useful discussions.
This study has in part been supported by Grant-in-Aid
for Scientific Research from the Ministry of Education,
Culture, Sports, Science and Technology of Japan.
The computations were performed
at the Supercomputer Center of
Institute for Solid State Physics,
and the Computer Center in the University
of Tokyo.
\par

\end{document}